\documentclass{article}
\usepackage[T1]{fontenc} % add special characters (e.g., umlaute)
\usepackage[utf8]{inputenc} % set utf-8 as default input encoding
\usepackage{ismir,amsmath,cite,url}
\usepackage{graphicx}
\usepackage{multirow}
\usepackage{color}
\usepackage{booktabs}
\usepackage{array}
\usepackage{bbding}

\usepackage[firstpage]{draftwatermark}
\definecolor{lightgray}{rgb}{0.9,0.9,0.9}
\definecolor{darkgray}{rgb}{0.4,0.4,0.4}
\SetWatermarkFontSize{12pt}
\SetWatermarkScale{1.1}
\SetWatermarkAngle{90}
\SetWatermarkHorCenter{202mm}
\SetWatermarkVerCenter{170mm}
\SetWatermarkColor{darkgray}
\usepackage{arydshln}
\usepackage{tikz}
\def\lbd{}	% Flag to use correct LBD settings in the paper, please do not modify this line

\definecolor{attcolor}{rgb}{0. , 0.5 , 0.9}
\definecolor{convcolor}{rgb}{1 , 0.7 , 0}
\definecolor{mlpcolor}{rgb}{0.9 , 0.3 , 0.4}
\definecolor{othercolor}{rgb}{0.8 , 0.4 , 0.9}
\definecolor{attconvcolor}{rgb}{0. , 0.6 , 0.1}
\definecolor{wihtecolor}{rgb}{1 , 1. , 1.}
\definecolor{blackcolor}{rgb}{0, 0. , 0.}
\definecolor{RowColor}{rgb}{0.97, 0.97, 1}

\newcommand{\attshape}{\raisebox{0.5pt}{\tikz\fill[attcolor] (0,0) circle (.8ex);}}
\newcommand{\convshape}{\raisebox{0.5pt}{\tikz\fill[convcolor] (0,0) circle (.8ex);}}
\newcommand{\mlpshape}{\raisebox{0.5pt}{\tikz\fill[mlpcolor] (0,0) circle (.8ex);}}
\newcommand{\convattshape}{\raisebox{0.5pt}{\tikz\fill[attconvcolor](0,0) circle (.8ex);}}
\newcommand{\othershape}{\raisebox{0.5pt}{\tikz\fill[othercolor] (0,0) circle (.8ex);}}

\title{MAP-Music2Vec: A Simple and Effective Baseline for Self-Supervised Music Audio Representation Learning}

\multauthor
{\fontsize{11}{12}\selectfont\vspace{-0.7mm}
Yizhi Li$^{1*}$\thanks{*\quad The authors contributed equally to this work.} \hspace{1cm} 
Ruibin Yuan$^{2,4*}$ \hspace{1cm} 
Ge Zhang$^{2,5*}$ \hspace{1cm}
Yinghao Ma$^{3*}$
} 
{ \fontsize{11}{12}\selectfont\vspace{-0.7mm} \bfseries{
    Chenghua Lin$^{1\dag}$\thanks{\dag\quad Corresponding authors.} \hspace{0.6cm} 
    Xingran Chen$^5$ \hspace{0.6cm} 
    Anton Ragni$^1$ \hspace{0.6cm}
    Hanzhi Yin$^4$ \hspace{0.6cm}
    Zhijie Hu$^6$}\\
\fontsize{11}{12}\selectfont\vspace{-0.7mm}\bfseries{ Haoyu He$^7$ \hspace{0.6cm}
    Emmanouil Benetos$^3$ \hspace{0.6cm}
    Norbert Gyenge$^1$ \hspace{0.6cm}
    Ruibo Liu$^8$ \hspace{0.6cm}
    Jie Fu$^{2\dag}$
    }\\
\fontsize{8}{9}\selectfont\vspace{-1.7mm}
$^1$Department of Computer Science, University of Sheffield, UK\\\fontsize{8}{9}\selectfont\vspace{-1.7mm}
$^2$Beijing Academy of Artificial Intelligence, China\\\fontsize{8}{9}\selectfont\vspace{-1.7mm}
$^3$Centre for Digital Music, Queen Mary University of London, UK\\\fontsize{8}{9}\selectfont\vspace{-1.7mm}
$^4$School of Music, Carnegie Mellon University, PA, USA\\\fontsize{8}{9}\selectfont\vspace{-1.7mm}
$^5$University of Michigan Ann Arbor, USA\\\fontsize{8}{9}\selectfont\vspace{-1.7mm}
$^6$HSBC Business School, Peking University, China\\\fontsize{8}{9}\selectfont\vspace{-1.7mm}
$^7$University of Tübingen \& MPI-IS, Germany\\\fontsize{8}{9}\selectfont\vspace{-1.7mm}
$^8$Dartmouth College, NH, USA\\
{\fontsize{9}{10}\selectfont\vspace{-1mm}
\tt\small \{yizhi.li, c.lin\}@sheffield.ac.uk, yinghao.ma@qmul.ac.uk, fujie@baai.ac.cn}
\vspace{-5mm}
}
\def\authorname{F. Author, S. Author, and T. Author}
\begin{document}
\maketitle
\begin{abstract}
\vspace{-0.8mm}
The deep learning community has witnessed an exponentially growing interest in self-supervised learning (SSL). 
However, it still remains unexplored how to build a framework for learning useful representations of raw music waveforms in a self-supervised manner. 
In this work, we design Music2Vec, a framework exploring different SSL algorithmic components and tricks for music audio recordings. 
Our model achieves comparable results to the state-of-the-art (SOTA) music SSL model Jukebox, despite being significantly smaller with less than 2\% of parameters of the latter.
The model will be released on Huggingface\footnote{Please refer to: https://huggingface.co/m-a-p/music2vec-v1}.

\vspace{-0.8mm}
\end{abstract}
\section{Introduction}
\label{sec:introduction}
\vspace{-0.8mm}
SSL has been proven effective for extracting features from raw music waveforms \cite{dhariwal2020jukebox, castellon2021codified}. 
Unfortunately, existing models (e.g., Jukebox \cite{dhariwal2020jukebox}) are prohibitively expensive to finetune and extend to different applications\footnote{Jukebox needs over 10GB to store activations.\vspace{-5mm}}. 
As an effort to obtain the computationally affordable baseline, we design and train Music2Vec. 
It mainly follows the design principles proposed in data2vec \cite{baevski2022data2vec}. 
Our key contributions are as follows: (1) developing music2vec, an open source self-supervised system for raw music files with single-GPU trainable size (about 90M parameters); (2) demonstrating that the model achieves comparable results to Jukebox on multiple music information retrieval tasks.

\vspace{-0.8mm}

\section{Method}

\begin{figure}[!t]
  \centering
  \includegraphics[width=0.75\linewidth]{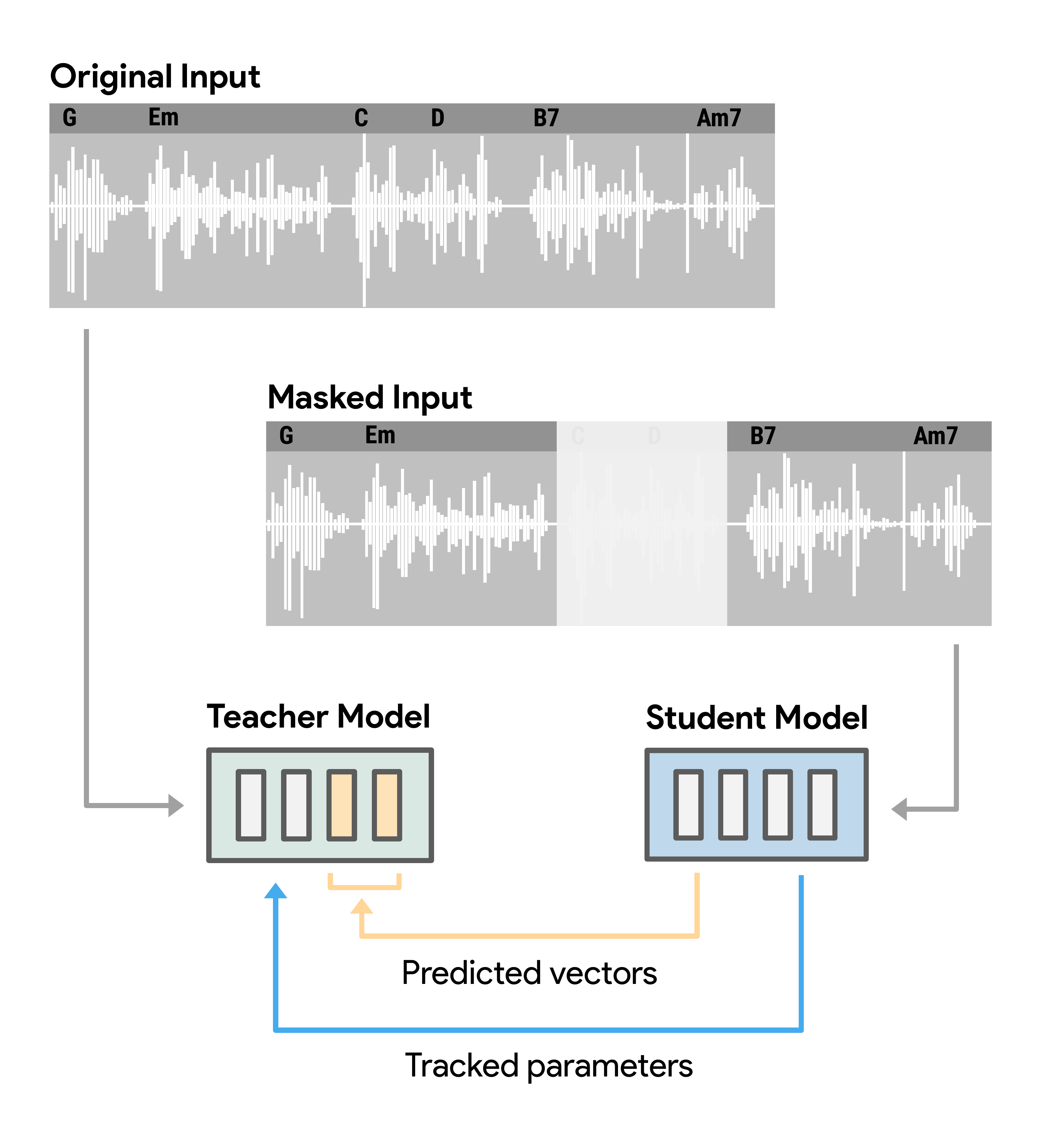}
  \vspace{-5.5mm}
  \caption{
    Music2Vec Framework. During pre-training, the student model aims to reconstruct the masked music audio by taking the contextualised representations provided by teacher model as prediction target.
  }\label{fig:overview}
  \vspace{-4.5mm}
\end{figure}

\vspace{-0.8mm}
We follow the standard pretraining protocols of data2vec~\cite{baevski2022data2vec} with the fairseq framework \cite{ott2019fairseq}, and further release our computationally affordable models. 
Data2vec claims a unified SSL framework for either speech, NLP, or computer vision following the design of BYOL \cite{grill2020bootstrap}, which is illustrated in Fig.~\ref{fig:overview}. 
The teacher model and student model share the same architecture, and the parameters of the teacher model are updated according to the exponential moving average of the student \cite{baevski2022data2vec}. 
The student model takes the partially masked input and is asked to predict the average of top-$K$ layer outputs of the Transformer in the teacher model. 
In contrast, the teacher model takes the unmasked input and provides contextual prediction targets in the pre-training.
% updates its parameters according to an exponential moving average of the student weights \cite{baevski2022data2vec}. 

\begin{table*}[bt]
\centering

\scalebox{0.85}{

\begin{tabular}{lllccccccc}
\hline
\multicolumn{3}{c}{\multirow{2}{*}{\textbf{Approach}}} & \multicolumn{2}{c}{\textbf{Tags ({\texttt{MTT}\cite{law2009evaluation}})}} & \textbf{Genre (\texttt{GTZAN}\cite{tzanetakis2002musical})} & \textbf{Key(\texttt{GS}\cite{knees2015two})} & \multicolumn{2}{c}{\textbf{Emotion ({\texttt{EMO}\cite{soleymani20131000}})}} & \multicolumn{1}{c}{\multirow{2}{*}{\textbf{Average}}} \\ 
% \cline{4-8} 
& & \multicolumn{1}{c}{}& \textbf{AUC} & \textbf{AP} & \textbf{Accuracy}& \textbf{Score}  & \textbf{{R2}{\tiny Arousal}}  & \textbf{R2{\tiny Valence}} & \\ 
\hline
% ---------------------------------------
\multicolumn{1}{c}{\multirow{4}{*}{\rotatebox[]{90}{\textbf{Baselines}}}} & \multicolumn{2}{c}{CHOI \cite{choi2017transfer}} &89.7 &36.4 &75.9 & 13.1 & 67.3 & 43.4 & 51.9 \\
& \multicolumn{2}{c}{MUSICNN \cite{pons2019musicnn}} & 90.6 &38.3 & 79.0 & 12.8 &  70.3 &  46.6 & 53.7 \\
& \multicolumn{2}{c}{CLMR \cite{spijkervet2021contrastive}} & 89.4 &36.1 & 68.6 & 14.9 &  67.8 &  45.8 &  50.8 \\
& \multicolumn{2}{c}{Jukebox \cite{dhariwal2020jukebox}} &\underline{91.5} &\underline{41.4} & \underline{79.7} & \underline{66.7} & \underline{72.1} & \underline{61.7} & \underline{69.9} \\
% ---------------------------------------
\midrule

\multicolumn{1}{c}{\multirow{11}{*}{\rotatebox[]{90}{\textbf{Music2Vec}}}} & \multicolumn{2}{l}{Starting Setting} & 88.2 & 34.1 & 61.7 & 32.1\textsuperscript{$\diamondsuit$}  & 66.2 & 45.8 & 54.7 \\
% \hdashline
% \cdashline{2-10}
\cline{2-10}
& \multicolumn{1}{l}{\multirow{3}{*}{\shortstack[l]{Length\\ Crop}}}  & \attshape{}5s & \fbox{89.5} & 35.9 & \fbox{76.6}& 50.1\textsuperscript{$\diamondsuit$} & \fbox{69.4} & \fbox{57.4} & \fbox{63.2} \\
&  & \attshape{}10s & 89.0 & \fbox{36.0} & 70.3 & 27.4 & 62.7 & 46.1 & 55.3 \\
&  & \attshape{}15s & 88.3 & 34.1 & 65.9 & 38.1\textsuperscript{$\diamondsuit$} & 60.1 & 43.6 & 55.0 \\
% \hdashline
\cdashline{4-10}
& \multicolumn{1}{l}{\multirow{2}{*}{\shortstack[l]{ Mask\\ Span}}}  & \convshape{}5 & 87.0\textsuperscript{$\diamondsuit$} & 32.2\textsuperscript{$\diamondsuit$} & 59.3 & 29.5 & 50.3\textsuperscript{$\diamondsuit$} & 24.7\textsuperscript{$\diamondsuit$} & 47.2 \\
& & \convshape{}15 & 87.8 & 33.3 & 65.2 & 41.9\textsuperscript{$\diamondsuit$} & 55.0 & 36.9 & 53.4 \\
% \hdashline
\cdashline{4-10}
 & \multicolumn{1}{l}{\multirow{3}{*}{\shortstack[l]{Mask\\ Prob}}}  & \mlpshape{} 50\% & 87.7 & 33.2 & 62.8 & 43.6\textsuperscript{$\diamondsuit$} & 54.8 & 37.6 & 53.2 \\
& & \mlpshape{} 70\% & 87.2 & 32.4 & 60.7 & 35.3\textsuperscript{$\diamondsuit$} & 55.3\textsuperscript{$\diamondsuit$}  & 36.0\textsuperscript{$\diamondsuit$}  & 51.2 \\
& & \mlpshape{} 80\% & 87.5 & 32.7 & 60.0\textsuperscript{$\diamondsuit$} & 34.6\textsuperscript{$\diamondsuit$} & 50.7 & 40.4 & 51.0 \\
% \hdashline
\cdashline{4-10}
% \multicolumn{1}{c}{\multirow{3}{*}{\shortstack[c]{Target\\ Layers}}}  & 6 & - & - & - & - & - & - & - \\
% & 10 & - & - & - & - & - & - & - \\
% & 12 & 88.8 & 34.5 & 65.2 & 50.8\textsuperscript{$\diamondsuit$} & 67.4 & 43.8 & - \\
% \cdashline{2-9}
& \multicolumn{1}{l}{\multirow{1}{*}{\shortstack[l]{Target}}}  
& {\convattshape}{}Top-12 & 88.8 & 34.5 & 65.2 & \fbox{50.8}\textsuperscript{$\diamondsuit$} & 67.4 & 43.8 & 58.4 \\
\cdashline{4-10}
& \multicolumn{1}{l}{\multirow{1}{*}{\shortstack[l]{Step}}}  & \othershape{}800K & 87.6\textsuperscript{$\diamondsuit$} & 33.2\textsuperscript{$\diamondsuit$} & 60.3 & 44.9\textsuperscript{$\diamondsuit$} & 54.8\textsuperscript{$\diamondsuit$}  & 40.8\textsuperscript{$\diamondsuit$}  & 53.6 \\
% \hdashline
% \cdashline{2-10}
\bottomrule
\end{tabular}
}
\caption{
Overall Results of Pre-trained Models. We report the results of Music2Vec trained with controlled variables derived from the speech data2vec setting \cite{baevski2022data2vec}. 
% \textcolor{blue}{
Underline and square box indicate the best overall performance and the best setting of Music2Vec, respectively.
% }
%The \underline{general best} performances and \fbox{Music2Vec best} settings are emphasised.
$\diamondsuit$ indicates the results are produced by the convolutional feature extractor representations.
We use dots with different colors to present different hyperparameters: \textcolor{attcolor}{length crop}, \textcolor{convcolor}{mask span}, \textcolor{mlpcolor}{mask prob}, \textcolor{attconvcolor}{target}, and \textcolor{othercolor}{step}. Results of baselines are taken from JukeMIR \cite{castellon2021codified} and datasets for different tasks are given in brackets.
}\label{tb:result}
\vspace{-1mm}
\end{table*}

We directly apply the Data2Vec base model, which encodes audio recordings using a multi-layer 1-D CNN feature extractor mapping 16 kHz waveform to 50 Hz representations~\cite{baevski2020wav2vec}, and further input the encoded tokens into a 12-layer Transformer Blocks with $H=768$ hidden dimension (with $4 \times H$ feed-forward inner-dimension). 
Since starting with pre-trained speech models can barely benefit music representation learning \cite{ragano2022learning}, we instead train the randomly initialised base model from scratch to verify its effectiveness on modelling music audio recordings.

We collect around 130k hours of music audio files from the Internet and use a 1k hours subset that contain 30s long wave files to train our model. 
All Music2Vec models are trained for 400k steps with 8 $\times$ NVIDIA A100-40GB GPUs. 
Training with eight GPUs takes around 6 days, i.e., about $48$ days with only one A100 GPU.

\vspace{-0.8mm}
\section{Experiments}{
\vspace{-0.8mm}
\subsection{Dataset and Evaluation}
\vspace{-0.8mm}
We follow the probing evaluation setting of JukeMIR \cite{castellon2021codified} to verify the music modelling performance of our models. 
Specifically, we report results on a comprehensive set of music information retrieval tasks, including multi-label \textbf{tagging}, multi-class \textbf{genre classification}, multi-class \textbf{key detection}, as well as a regression task \textbf{emotion recognition}. 
% The metric $MTT_{AUC}$ denotes the area under the receiver operating characteristic curve on MTT dataset, and second metrics $MTT_{AP}$ denotes the average precision on MTT dataset. 
% The second metric is different with previous methods but is used in the SOTA pretrained model for music, so we remain it as the same.
Following the evaluation setting \cite{castellon2021codified} for the SOTA pre-trained model, AUC (the area under the receiver operating characteristic curve) is regarded as the main metric of tagging to select checkpoints, and the marco average of arousal and valence R2 decides for emotion recognition.
\vspace{-0.8mm}

\subsection{Pre-train Settings}

\vspace{-0.8mm}
Adapting from the data2vec model on auditory signals\footnote{We use audio files with 30s length, mask span length 10, mask probability 65\%, target top-$8$, and training step 400K as the starting setting. The results is shown as the starting setting in the table.}, we conduct parameter searching and correlation analysis for Music2Vec pretraining, including the recording length, the mask strategy, and the learning target layers.

\textbf{First,} we use \textbf{audio length cropping} to shorten music excerpts, since longer sequences are more difficult for modelling.
Note that the combined audio file length in a batch is not altered and the hardware environment remains the same, which makes a single training batch contains larger number of music samples when cropping the clips.

\textbf{Second,} we revise the mask strategy by changing \textbf{mask span length} and \textbf{mask token probability}. 
Mask token probability is the probability for each token to be chosen as the start of the span to be masked, and the length of which can also be adapted for different data modalities \cite{baevski2022data2vec}. 
% Mask span length is defined as the length of consequent tokens when masking. 

\textbf{Third,} we modify the \textbf{prediction target} provided by the teacher model. 
Our preliminary experiments illustrate that early layer representations generally perform well on key detection. 
% Therefore, we try to change the prediction target in Music2Vec from the average of the top-8 layer representations to all the 12 layers, so that the student model might learn from target representations that potentially preserve more musical key information. 
Therefore, we change the prediction target in Music2Vec from the average of the top-8 layer representations to all the 12 layers, so that the student model might benefit from the potentially preserved key information. }

\vspace{-0.8mm}
\section{Results and Conclusion}
\vspace{-0.8mm}
From Tab.~\ref{tb:result} we observe that the Music2Vec with the best setting (i.e., crop5s) achieves comparable results to Jukebox on music information retrieval tasks with less than 1/50 parameters of the latter. 
The audio file length is negatively correlated to the Music2Vec performance, which implies that modelling long sequence is still challenging.

Noticeably, the CNN representations sometimes outperform the Transformer layers, especially for key detection.
When including extra early Transformer layers to the prediction target, Music2Vec achieves performance gains in most tasks.
% Transformer layers from the teacher
This implies that with little or no contextualisation
%with less or even without contextualisation,
our model still manages to perform fairly well with local features (similar to bag-of-words).
% However, this also suggests that the model relies too much on local information and there is still a large room for improvement when taking longer context into consideration.
However, this also suggests that our model relies too much on local information and leaves a large room for improvement when taking long-range contextual information into consideration. 
Last but not least, we find that increasing the training steps, changing the mask span, or changing the mask probability does not give performance gain in most tasks.

In conclusion, we propose a training framework for music audio recording pre-training on large-scale data, which gives comparable performance to SOTA models. 
Our framework also has great potential for efficient fine-tuning and model distillation, which we leave for future work.

\section{Acknowledgements}

This paper is a tribute to our talented friend Anqiao Yang, for his friendship and valuable advice to this work. 
Yizhi Li is fully funded by an industrial PhD studentship (Grant number: 171362) from the University of Sheffield, UK. 
Yinghao Ma is a research student at the UKRI Centre for Doctoral Training in Artificial Intelligence and Music, supported by UK Research and Innovation [grant number EP/S022694/1]. 
This work is supported by the National Key R\&D Program of China (2020AAA0105200). 
We acknowledge IT Services at The University of Sheffield for the provision of services for High Performance Computing.
We would also like to express great appreciation for the suggestions from faculties Dr Chris Donahue, and Dr Roger Dannenberg, as well as the facility support from Mr. Yulong Zhang in the preliminary stage.

% \newpage
% For bibtex users:

\bibliography{ISMIRtemplate}
% For non bibtex users:
%\begin{thebibliography}{citations}
% \bibitem{Author:17}
% E.~Author and B.~Authour, ``The title of the conference paper,'' in {\em Proc.
% of the Int. Society for Music Information Retrieval Conf.}, (Suzhou, China),
% pp.~111--117, 2017.
%
% \bibitem{Someone:10}
% A.~Someone, B.~Someone, and C.~Someone, ``The title of the journal paper,''
%  {\em Journal of New Music Research}, vol.~A, pp.~111--222, September 2010.
%
% \bibitem{Person:20}
% O.~Person, {\em Title of the Book}.
% \newblock Montr\'{e}al, Canada: McGill-Queen's University Press, 2021.
%
% \bibitem{Person:09}
% F.~Person and S.~Person, ``Title of a chapter this book,'' in {\em A Book
% Containing Delightful Chapters} (A.~G. Editor, ed.), pp.~58--102, Tokyo,
% Japan: The Publisher, 2009.
%
%
%\end{thebibliography}
\end{document}